\pdfoutput=1
\documentclass[%
 reprint,
nofootinbib,
 amsmath,amssymb,
 aps, prd,
floatfix,
showkeys,
]{revtex4-1}

\usepackage{graphicx}
\usepackage{dcolumn}
\usepackage{bm}

\begin{document}

\preprint{}

\title{Dark Energy from Ratio Gravity}

\author{Jackie C.H. Liu}
 \email{chjliu@ust.hk}
\author{Yi Wang}%

\affiliation{%
 Department of Physics, The Hong Kong University of Science and Technology, Clear Water Bay, Kowloon, Hong Kong, P.R.China
}%

\date{\today}

\begin{abstract}
The theory of Ratio Gravity (RG) proposes that the curvature of 3+1 spacetime  originates from a deformation of a Cross Ratio, where similar mathematical structure to general relativity emerges.  This paper studies RG using the framework of the Newman-Penrose spin formalism. After proposing the general formalism, we move on to study a homogeneous and isotropic universe with RG. It is noted that the theory contains a component of dynamical dark energy with novel equation of state. 
\end{abstract}

\keywords{Cosmology, Gravity, General Relativity}
\maketitle

\section{\label{sec:level1}INTRODUCTION}
 
Dark energy (DE) is one of the major mystery in cosmology
\cite{Riess:1998cb, Perlmutter:1998np}. A great number of dark energy models are built since its discovery \cite{Li:2011sd}.

Cross ratio is also known as the anharmonic ratio. It has been used in theoretical physics such as in conformal field theory \cite{CFT} and scattering amplitudes \cite{Huang2016}. In the present work, we study a theory of gravity known as Ratio Gravity (RG) \cite{Liu2016} as the first principle of gravity, and explore its explanation of DE. A short introduction of the previous work is given in section 2.

We reformulate the RG theory by using the Newman Penrose (NP) formalism of general relativity \cite{NP1962}.  The correspondence between the NP formalism and RG theory is described in section 3, which demonstrates the connection between general relativity and RG. This shows evidence that the Cross Ratio may be the underlying mathematical structure of spacetime and gravity.

In section 4, we make use of the study of cosmological structure by NP formalism, then solve the equations of RG theory for a simple model.  The Friedmann equation and the equation of state of DE, $w(z)$, in RG are obtained.

The results show that DE is dynamical with the equation of state approximately -1 in the present era, which is consistent to the widely accepted $\Lambda$CDM model.  Furthermore, $w(z)$ approaches zero as $z$ approaches high z regime. This is an interesting feature, since recent observations show some evidences that dark energy may have similar dynamics \cite{Zhao:2017cud}. An observational best-fit polynomial plot for $w(z)$, with interesting $w(z)=-1$ crossing%
\footnote{The crossing of $w(z)$ to the line of $w=-1$ has been studied in the literature of dynamical dark energy. See \cite{Feng:2004ad}, and \cite{Copeland:2006wr, Cai:2009zp, Bamba2012, Wang:2016och} for reviews.}, is compared to the plot of $w(z)$ of RG, with good agreement. The expansion history of the simple model shows good match to observation.

Furthermore, the proportional constant of $\Omega_{b}$ and $\Omega_{M}$ is originated from the context of RG in this simple model. Finally, we discuss the potential extensions of the theory.

\section{OVERVIEW OF THE PREVIOUS WORK}

Here we outline the foundation of the previous work about Ratio Gravity \cite{Liu2016}, which is needed for the formulation of this paper.  Throughout this paper, we will use the definitions and conventions of the spinor formalism of general relativity developed by Penrose and Newman (\cite{Penrose1960}, \cite{NP1962}).  An intensive introduction for the spinor formalism is covered in chapter 13 of \cite{wald1984}.

Cross Ratio (\textit{Ratio}), defined as
\begin{equation}
 (z_1, z_2; z_3, z_4) = \frac{(z_3-z_1)(z_4-z_2)}{(z_3-z_2)(z_4-z_1)}  \label{eq:ratioEqn}, 
\end{equation}
is postulated to play a fundamental role for nature of gravity. In the context of gravity, the principle can be stated as: 
\textit{Gravity is described by the general deformation of the same cross-ratio.}   Such Cross Ratio consists of many equivalent representations - the general deformation.  In RG theory, the Cross Ratio is not used as a technical tool as in the previous application of Cross Ratio in theoretical physics \cite{CFT, Huang2016}. Instead, it is used to describe the mathematical structure of gravity.

The arbitrariness of the same cross ratio (L.H.S. of equation \ref{eq:ratioEqn}) leads to 4 degrees of freedom because of only 4 free parameters for 3 moving poles from R.H.S. of equation (\ref{eq:ratioEqn}) over Riemann sphere at an arbitrary reference point $z_1$.  One of the representation (or called realization) for cross ratio described in \cite{Yoshida} is hypergeometric differential equation $E(a,b,c)$, which has solution of usual hypergeometric series $F(a,b,c;z)$.  The theory proposed to represent the Ratio by an integratable system with 3 regular singular points, because the hypergeometric differential equation with 3 regular singular points is the corresponding representation of the cross ratio \cite{Yoshida}.  The following is a simple illustration for the correspondence of Ratio and differential equation
\begin{equation}
(z_1, z_2; z_3, z_4) = \frac{(z_3-z_1)(z_4-z_2)}{(z_3-z_2)(z_4-z_1)}
\nonumber
\end{equation}\vspace{-0.4cm}
\begin{equation}
	\updownarrow \nonumber
\end{equation}\vspace{-0.7cm}
\begin{equation}\nonumber
\mbox{Hypergeometric differential equation (3 regular singular poles)}.
\end{equation}

The main idea of Ratio Gravity is that the transformation of the Ratio manifests the variety of representations such that each representation is correspondent to related physical event/observation.

In 1960, Penrose \cite{Penrose1960} proposed spinor General Relativity formalism, which later was developed to well known NP formalism \cite{NP1962}.  It describes 3+1 spacetime structure by spinor formalism.  The key equations to describe curvature of spacetime are following two curvature spinor equations\footnote{Following the convention used in  \cite{Penrose1960}, \cite{NP1962}, capital letter indexes are used for spinor, while small letter indexes are used for dyad.}:
\begin{equation}
  (\partial_{AF'}\partial_{B}^{F'}+\partial_{BF'}\partial_{A}^{F'})\xi_{Q}=\chi_{ABQC}\xi^{C}				\label{eq:chiSpinor}
\end{equation} and 
\begin{equation}
  (\partial_{EC'}\partial^{E}_{D'}+\partial_{ED'}\partial^{E}_{C'})\xi_{Q}=\Phi_{QAC'D'}\xi^{A},				\label{eq:phiSpinor}
\end{equation}
where $\partial_{AF'}$ is spinor covariant derivative, and $\chi_{ABQC}$ and $\Phi_{QAC'D'}$ are called curvature spinors.  The curvature spinors require the following spinorial Bianchi identities to describe the dynamics of the spacetime structure:
\begin{equation}
  \partial^{D}_{G'}\chi_{ABCD}=\partial_{C}^{H'}\Phi_{ABG'H'}				\label{eq:spinorBianchi}.
\end{equation}

It is clear that we should find the connection between the Cross Ratio and the equations of spacetime structure described by equations (\ref{eq:chiSpinor}, \ref{eq:phiSpinor}, and \ref{eq:spinorBianchi}).

For the realization of gravity in RG theory,  Ratio Gravity postulates that gravitational curvature spinors (which are described by the Newman and Penrose spinor formalism of general relativity \cite{NP1962}) are originated by two sets of   underlying differential equation from Ratio defined as (\ref{eq:JMOp}): 
\textit{$J$,$M$ operators}, called as \textit{curvature operators}.
They generate 2 sets of 2-by-2 traceless matrices ($J_{ac}$, $M_{a\prime c\prime }$, or simply $J$,$M$) as \textit{curvature matrices} by operating on a 2-by-2 invertible matrix $Y$ (\textit{fundamental solution matrix})~\cite{Liu2016}:
\begin{equation}
J_{ac}(Y)=J_{ac}Y, M_{a\prime c\prime }(Y)=M_{a\prime c\prime }Y  				\label{eq:JMOp}, 
\end{equation}
where
\begin{equation}
J_{ac}(Y):=\epsilon^{b\prime d\prime }(B_{ab\prime } D_{cd\prime }-B_{cd\prime } D_{ab\prime })Y \label{eq:JOpDef},
\end{equation}
\begin{equation}
M_{b\prime d\prime }(Y):=\epsilon^{ac}(B_{ab\prime } D_{cd\prime }-B_{cd\prime } D_{ab\prime })Y \label{eq:MOpDef}
\end{equation}
($a,c $ and $a\prime,c\prime$ are spinor(dyad) indexes from 0 to 1 and from $0\prime$ to $1\prime$ respectively; $\epsilon^{b\prime d\prime }$ is the usual skew spinor metric for spinor formalism that is used to raise or lower the spinor indexes\footnote{In this paper, we use the convention $\epsilon_{ab}$ defined as $
\left(
\begin{array}{cc}
 0 & -1 \\
 1 & 0 \\
\end{array}
\right)$
}).  $B_{ab\prime }$ are four 2-by-2 matrices indexed by $ab\prime$;  $D_{cd\prime}$ are four 2-by-2 matrix-differential operators indexed by $cd\prime$, which obeys Leibniz's rule for derivation: 
\begin{equation}
D_{ab\prime }( XY )=(D_{ab\prime }X)Y+X(D_{ab\prime }Y),
\end{equation}
where $X$, $Y$ are differentiable 2-by-2 matrices.  The matrix differential equation corresponding to Ratio is defined as:
\begin{equation}
D_{ab\prime }Y=B_{ab\prime }Y.							\label{eq:YEqn}
\end{equation}
This is called \textit{$Y$equation}.  We will see in the next section that  $B_{ab\prime }$ matrices have direct physical meaning related to the curvature structure of spacetime, and $D_{ab\prime }$ operators are related to the metric.  Therefore, the $Y$ equation describes how metric and curvature are related.

Equation (\ref{eq:JMOp}) is defined because of two reasons: first, it has the similar algebraic form as curvature spinor equations (\ref{eq:chiSpinor}) and (\ref{eq:phiSpinor}) (after few spinor algebraic manipulations and the use of symmetry of the curvature spinors\footnote{$\chi_{ABCD}=\chi_{CDAB}$, $\Phi_{ABG'H'}=\Phi^{*}_{G'H'AB}$}); secondly, it is related to the integratability condition of underlying differential equation (Ratio representation in this case): 
\begin{equation}
 d_j B_i - d_i B_j = B_j B_i - B_i B_j   				\label{eq:intCondEqn} 
\end{equation}
(where $d_i$ is derivative operator of the differential equation), that is closely connected to the property of regular singular points of Ratio \cite{Cassidy}.

The spinor formalism defines spinor over complex number, while the equations of RG theory are written by spinor-indexed 2-by-2 matrices or 2-by-2 matrix operators, so the basic object is in 2-by-2 matrix structure over complex number.

For the second part of spinor General Relativity - spinorial Bianchi identities (\ref{eq:spinorBianchi}), RG leverages the Galois differential theory as the tool for deforming the Ratio such that the dynamics of the curvature operators $J, M$ (as \textit{Gal operators} are defined below) is manifested similarly as spinorial Bianchi identities.  In particular, Ratio needs an extension of the theory of Galois differential theory to deform the Ratio (parameterized Picard-Vessiot theory, PPV theory for short) \cite{Cassidy}. 

There are few key points for PPV theory:
\begin{itemize}
	\item PPV deformation can be used for linear differential matrix equation that the deformation freedom extends to parameter spaces of the differential equation.  We can simply think of the solution space of the differential equation forms a vector space, and the deformation is equivalent to the transformation of the solution vector and the system over the parameters space.
\end{itemize}
\begin{itemize}
	\item Gal operator (usually labelled as $\sigma$) is defined that it commutes with differential operators ($D$) when operating on Y: $D\cdot{\sigma}(Y)=\sigma\cdot{D}(Y)$.  This property is called \textit{Gal-D-commuting property} in this paper.
\end{itemize}
\begin{itemize}
	\item 	The Galois differential group is important that it contains the characteristics of the underlying differential equation.  In fact, Galois differential theory was developed in order to solve the differential equation by transforming to less complicated system.
\end{itemize}

As an example, one may write differential equation, e.g. $y\prime\prime-y=0$, in matrix form (i.e. $Y$ is the fundamental solution matrix) and the Galois group is $SL_{2}$.  By defining a Gal operator that satisfies $\sigma(Y)=YC$, where $C$ is $SL_{2}$ constant matrix for $D$ operator (i.e. $D(C) = 0$), then,
$D\cdot{\sigma}(Y)=\sigma\cdot{D}(Y)$ implies 
\begin{equation}
	DZ=BZ, \sigma(Z)=ZC,
\end{equation}
where $Z:=\sigma(Y)$ (In fact, $C$ is diagonal and in $SL_{2}$).  So the new differential equation and solution are obtained.

Realizing the form of spinor Bianchi identities (\ref{eq:spinorBianchi}) and the definition of $J, M$ curvature operators (\ref{eq:JMOp}, \ref{eq:JOpDef}, \ref{eq:MOpDef}), it is clear and natural to define the Bianchi equation for the version of Ratio Gravity as \cite{Liu2016}:
\begin{equation}
	D^{b}_{d\prime}J_{ab}(Y)=D_{a}^{c\prime}M_{c\prime d\prime}(Y).	\label{eq:DJDM0}
\end{equation}  In fact, it is important to point out that the construction of (\ref{eq:DJDM0}) can be justified by contracting the commutation equation of Gal operators ($J$, $M$) and differential operators ($D_{a}^{c\prime}$,$D^{b}_{d\prime}$) up to spinor indexes $b$ and $c'$ accordingly.  Furthermore, matrices $J$, $M$ are identified as the algebras of Galois differential group of Ratio (Hypergeometric differential equation representation) $SL_{2}$ \cite{Beukers}.

As a summary, RG theory leverages the curvature operators and the associated Bianchi equation to establish the connection to the spacetime structure.

\section{USE OF NEWMAN PENROSE SPINOR FORMALISM}

In this section, the well-known Newman Penrose formalism \cite{NP1962} is used.  We explain how RG theory establishes the connection to NP formalism such that the mathematical structure like Bianchi identities emerges from RG theory.

In NP formalism, there are two parts of equations: NP equations and the Bianchi identities.  The NP equations provides geometric properties (through spin coefficients $\Gamma_{abcd\prime}$\footnote{Spin coefficients $\Gamma_{abcd\prime}$ defined in NP formalism as: \\ $\Gamma_{abcd\prime}$=
$
\begin{array}{|c|ccc|}\cline{1-4}
 \text{cd}' \slash \text{ab} & \text{00} & | \text{01 or 10}| & 11 \\\cline{1-4}
 00' & \kappa  & \epsilon  & \pi  \\
 10' & \rho  & \alpha  & \lambda  \\
 01' & \sigma  & \beta  & \mu  \\
 11' & \tau  & \gamma  & \nu  \\\cline{1-4}
\end{array}
$
}) from curvature spinors, which is postulated to be originated by curvature matrices in dyad:
\small
\begin{multline}
-J_{\text{ab}}\epsilon =\Gamma _{\text{ac$\prime $}} 
\text{ }^{d\prime}\epsilon \bar{\Gamma }_{\text{d$\prime $b}}\epsilon^{c\prime}
+\Gamma _{\text{bc$\prime $}} \text{ }^{d\prime}\epsilon \bar{\Gamma }_{\text{d$\prime $a}}\epsilon^{c\prime}+\\
\epsilon^{d\prime c\prime} (\Gamma _{\text{ad$\prime $}}\epsilon \Gamma _{\text{bc$\prime $}}-\Gamma _{\text{bc$\prime $}}\epsilon \Gamma _{\text{ad$\prime $}}+\\
\partial _{\text{bc$\prime $}}\Gamma _{\text{ad$\prime $}}-\partial _{\text{ad$\prime $}} \Gamma _{\text{bc$\prime $}}+\epsilon^{wh} (\Gamma _{\text{habc$\prime $}} \Gamma _{\text{wd$\prime $}}-\Gamma _{\text{hbad$\prime $}} \Gamma _{\text{wc$\prime $}})) \label{eq:RGNPEqnJ},
\end{multline}
\begin{multline}
-M_{\text{c$\prime $d$\prime $}}\epsilon =\Gamma _{\text{bd$\prime $}} 
\text{ }^{a}\epsilon \Gamma_{ac\prime}\epsilon^{b}
+\Gamma _{\text{bc$\prime $}} \text{ }^{a}\epsilon \Gamma_{ad\prime}\epsilon^{b}+\\
\epsilon^{ab}
(\Gamma _{\text{ad$\prime $}}\epsilon \Gamma _{\text{bc$\prime $}}-\Gamma _{\text{bc$\prime $}}\epsilon \Gamma _{\text{ad$\prime $}}+\\
\partial _{\text{bc$\prime $}}\Gamma _{\text{ad$\prime $}}-\partial _{\text{ad$\prime $}} \Gamma _{\text{bc$\prime $}}+\epsilon^{w\prime h\prime}  (\Gamma _{\text{aw$\prime $}} \bar{\Gamma }_{\text{h$\prime $d$\prime $c$\prime $b}}-\Gamma _{\text{bw$\prime $}} \bar{\Gamma }_{\text{h$\prime $c$\prime $d$\prime $a}})),
	\label{eq:RGNPEqnM}
\end{multline}
\begin{eqnarray}
X_{\text{ab}}= -J_{\text{ab}} \epsilon,
\end{eqnarray}
\begin{eqnarray}
\Phi _{\text{c$\prime $d$\prime $}}{}^* = -{M}_{\text{c$\prime $d$\prime $}} \epsilon,
\end{eqnarray}
\normalsize
where \footnote{$\Gamma_{abcd\prime}: \Gamma_{cd\prime}$ as matrix with components indexes a,b; $\text{ }^{a}\epsilon \Gamma_{cd\prime}\epsilon^{b}$: a,b components of matrix $\epsilon \Gamma_{cd\prime} \epsilon$} $\epsilon$ is the spinor metric in its matrix form and ($\Phi_{c\prime d\prime}^{*}$,$X_{ac}$)\footnote{By traceless property of $J_{ac}$, $X_{ac}$ is symmetric matrix corresponding to curvature dyads $X_{abcd}$ in the NP formalism (same as $\Phi_{c\prime d\prime}^{*}$)}  are matrix forms of curvature dyad in the NP formalism. They are the counter part of the components of Riemann tensor in the metric formulation of GR, to represent the commutation relations of covariant
derivatives.  The equations (\ref{eq:RGNPEqnJ}) and (\ref{eq:RGNPEqnM}) defines \textit{NP equation in RG}.

The NP equation in RG and the original NP equations are different in the way that it is written.  Since original NP equations does not use matrix representation, so $\Gamma _{\text{bd$\prime $}} $ is written as $\Gamma _{\text{acbd$\prime $}} $.

For Bianchi identities in the NP formalism, it describes how changes of curvature dyad and spin coefficients are connected.  The Bianchi identities are indirectly originated by the spinor Bianchi equation \cite{Liu2016} with dyad as following:
\begin{equation}
	D^{b}_{d\prime}J_{ab}(Y)=D_{a}^{c\prime}M_{c\prime d\prime}(Y),	\label{eq:DJDM}
\end{equation}
while the Ratio ($Y$) equation for dyad $D$ differential operator is defined as
\begin{equation}
	D^{b}_{d\prime}Y={\partial }^{b}_{d\prime}Y+\left[{\Lambda }^{b}_{d\prime},Y\right]= B^{b}_{d\prime}Y \label{eq:DYBY},
\end{equation}
where ${\Lambda }^{b}_{d\prime}$ are four traceless 2-by-2 matrices and ${\partial }^{b}_{d\prime}$ are defined as identity 2-by-2 matrix multiplying the dyad differential operators\footnote{${\partial }_{bd\prime}:= \sigma^{\mu}_{bd\prime} \partial_\mu$, where $\partial_\mu$ is covariant derivative, and $\sigma^{\mu}_{bd\prime}$ is Hermitian matrix connecting metric tensor to metric dyad \cite{NP1962}.} in the NP formalism (note that the original dyad differential operators are spinor operators, not matrix operators).  The dyad differential operators contain the geometric structure variables (e.g. scaling factor in cosmology) of the spacetime metric.  $J$ and $M$ operators are the Gal operator discussed in section 2, so it is clear that they play fundamental role in describing the curvature.
The dyad derivatives of curvature matrices are described by Ratio \textit{context} ($B_{ab\prime}, \Lambda_{ab\prime}, Y$).  After computing the Bianchi identities explicitly from original spinor formalism to dyad formalism (the same procedure used by the NP formalism \cite{NP1962}, section II-Tetrad Calculus), we can connect the spin coefficients $\Gamma_{\text{hbfd$\prime $}}$ and Ratio (context) to construct \textit{Bianchi Constraints}:
\small
\begin{multline}
	\epsilon^{c\prime b\prime} \epsilon^{g\prime h\prime} \left(\mathcal{M}_{\text{c$\prime $g$\prime $}} \bar{\Gamma }_{\text{h$\prime $d$\prime $b$\prime $a}}+\mathcal{M}_{\text{g$\prime $d$\prime $}} \bar{\Gamma }_{\text{h$\prime $c$\prime $b$\prime $a}}\right)-\\
	\epsilon^{bf} \epsilon^{gh} \left(\mathcal{J}_{\text{ag}} \Gamma_{\text{hbfd$\prime $}}+\mathcal{J}_{\text{gb}} \Gamma _{\text{hafd$\prime $}}\right)\\
	=-[\Lambda^{b}_{\text{d$\prime $}},\mathcal{J}_{\text{ab}}]-\mathcal{J}_{\text{ab}} B^{b}_{\text{d$\prime $}}+\\
	[\Lambda^{c\prime}_a,\mathcal{M}_{\text{c$\prime $d$\prime $}}]+\mathcal{M}_{\text{c$\prime $d$\prime $}} B^{c\prime}_a ,
	\label{eq:BC}
\end{multline}
\normalsize
i.e. the geometrical construction in dyad as Bianchi identities is established as Bianchi constraints.

\begin{figure}[ht]
\includegraphics[width=0.40\paperwidth] {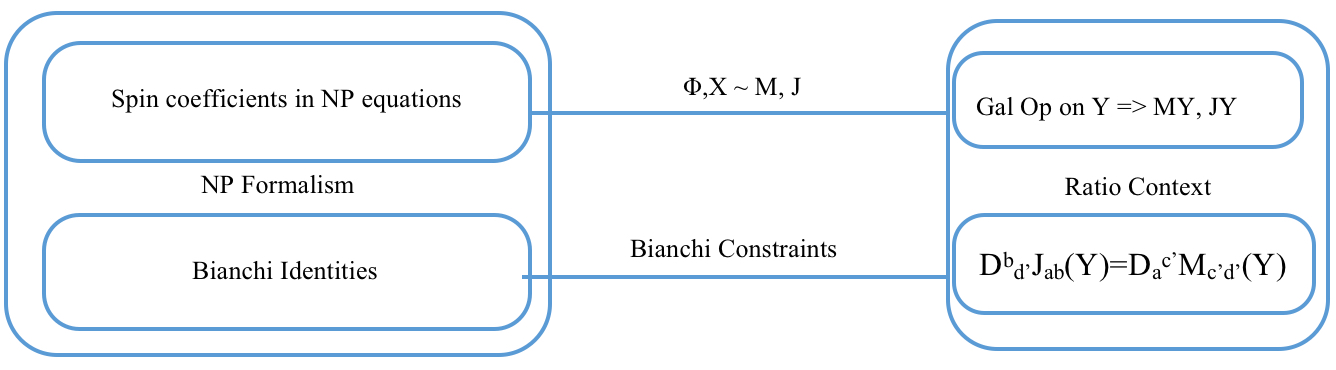}
\caption{\label{fig:RGDERatioNP} The correspondence of the NP formalism and the Ratio context.}
\end{figure}

Physically, Ratio describes how the geometry of spacetime by mean of spin coefficients is manifested.  It is even clearer if we notice Bianchi constraints are in fact algebraic.  The diagram [Fig.~\ref{fig:RGDERatioNP}] shows that RG equations connecting the related curvature dyads to spinor coefficients to Bianchi identities. The Bianchi constraints are constructed to map the Bianchi identities to RG dyads. So it forms a set of consistent equations, and, in principle, is solvable.

In short, Ratio provides dyads dynamics to establish a 3+1 spacetime.  It also provides the related differential operators for the RG to realize as curvature dyads, which are solved with Bianchi constraints.

In the last part of this section, we illustrate the correspondence of RG theory and general relativity in brief glance.  We start with equations of the NP formalism and show them lead to RG equations.  

First, we assume $J_{ab}$ and $M_{c\prime d\prime}$ operators generate $J_{ab}$ and $M_{c\prime d\prime}$ curvature matrices without requiring they are Gal operators for a moment, so we ``relax'' the constraint of being Gal operators.  The NP equations of RG theory (\ref{eq:RGNPEqnJ},\ref{eq:RGNPEqnM}) are exactly the same to NP equations of the NP formalism with direct substitution of curvature dyads to curvature matrices.  Therefore, we only need to show how Bianchi identities of the NP formalism lead to RG Bianchi equation (\ref{eq:DJDM}).  The Bianchi identity of the NP formalism consists of the term, $\partial^{b}_{d\prime}J_{ab}-\partial_{a}^{c\prime}M_{c\prime d\prime}$, which is the the left hand side of Bianchi constraints (\ref{eq:BC}) of RG equations.  By introducing an invertible 2 by 2 matrix $Y$, and simply using $Y$ equation (\ref{eq:YEqn}) as a constraint equation, RG Bianchi equation (\ref{eq:DJDM}) is recovered.  The use of both constraints (Bianchi constraints and $Y$ equation as constraint equation) is satisfied by the number of unknowns of RG context - all the components of $Y$, $B_{ab\prime}$, and traceless matrices $\Lambda_{ab\prime}$, total 4 $\times$ (1 + 4 + 3) = 32 unknowns.

In this brief discussion, it is required that we relax the constraints for $J_{ab}$ and $M_{c\prime d\prime}$ operators as defined in section 2, to illustrate the correspondence.  However, as stated in previous work \cite{Liu2016}, the Gal operators are naturally related to RG Bianchi equation.  So simply relaxing the constraints for $J_{ab}$ and $M_{c\prime d\prime}$ operators defined in section 2 is not correct approach in the context of RG theory.  Instead, other forms of Gal operators as the extension of the theory are perfectly valid under the framework of the RG theory.

\section{THE SOLUTION FOR THE COSMOLOGY IN HOMOGENEOUS METRIC}

To illustrate the framework described by RG in the NP formalism and its predictions, we apply it to construct cosmology with a simple solution. 
A widely used metric form for FRW cosmology is $ds^2=\mathit{a}(\eta)^2(d\eta ^2-dr^2-\mathcal{S}(\mathit{r})^2 (d\theta^2+\sin ^2\theta d\phi^2))$.  The dyad differential operators associated are
\small
\begin{multline}
	\partial _{00'}=\frac{\partial _{\mathit{r}}+\partial _{\eta }}{\mathit{a}(\eta )^2},\partial _{11'}=\frac{1}{2}\left(-\partial _{\mathit{r}}+\partial _{\eta }\right),\\
	\partial _{01'}=\frac{\partial _{\theta }+\frac{i\partial _{\phi }}{\sin (\theta )}}{\sqrt{2} \mathit{a}(\eta ) \mathcal{S}(\mathit{r})},\partial _{10'}=\frac{\partial _{\theta }-\frac{i\partial _{\phi }}{\sin (\theta )}}{\sqrt{2} \mathit{a}(\eta ) \mathcal{S}(\mathit{r})}
	\label{eq:dyadD}.
\end{multline}
\normalsize
The condition of the spin coefficients is constructed by the cosmological principles for homogeneous spacetime, we have spin coefficients conditions \cite{Stephani}:
\begin{equation}
 \kappa=\rho=\sigma=\epsilon + \epsilon^{*}=\tau+\beta+\alpha^{*}=0.
\end{equation}
And the dependency of $Y$ is limited to only $\eta$, $r$; we parameterize, as PPV-theory-sense, only to $D_{00\prime}$, $D_{11\prime}$.
A solution is found by a simple model after solving the overdetermined system:
\begin{enumerate}
	\item $Y$ equation (\ref{eq:YEqn}),
	\item RG Bianchi equation (\ref{eq:DJDM}),
	\item Gal-D-commuting property: \\ $[D_{ab\prime}, \hat{J_{cd}}]Y=0$,$[D_{ab\prime}, \hat{M_{c\prime d\prime}}]Y=0$,
	\item Bianchi constraints (\ref{eq:BC}),
	\item NP equation in RG (\ref{eq:RGNPEqnJ},\ref{eq:RGNPEqnM}),
\end{enumerate}	
which we need the spin coefficients condition ($\beta=\tau=\pi=\nu=\lambda=0$), and condition of two constants - ($\mu=\mu_{0}/2$, and $C_{00\prime}$ =scalar part of $B_{00\prime}$), as well as the condition of $Y$ matrix components: $Y_{00}, Y_{11} \ll Y_{10}$.  The first three equations (\#1 to \#3) are explained in previous work \cite{Liu2016}.  The $4^{th}$ and $5^{th}$ equations are developed in this paper because of the use of the NP formalism in dyad.  These additional two equations are needed when realization of the NP formalism in dyad, while, in the previous work, only spinor formalism is used.  As a result, the connection to metric thru the null tetrad by the NP formalism in dyad is clear.

The calculation is greatly simplified after realizing the zeros of spin coefficients in Bianchi constraints and homogeneous of $\Phi$ - curvature dyad - terms such that the Ratio context is restricted to be of a simpler form.  \textit{This is a simple solution so other cosmological solutions are certainly possible}.

As a result, the differential equations we need to describe the spin coefficients and curvature dyad are:
\begin{eqnarray}
\epsilon (\gamma + \gamma^{*})-\epsilon\prime /2+\gamma\prime / \mathit{a}^{2}=0,
\end{eqnarray}
\begin{eqnarray}
\Phi_{22}+2 \mu_{0}(\mu_{0}+2 Re(\gamma))=0,
\end{eqnarray}
\begin{eqnarray}
\epsilon^{*}= C_{00}/2- \Phi_{22} \prime/(2 \mathit{a}^{2} \Phi_{22}),
\end{eqnarray}
where prime stands for $\eta$ derivative, and $\Phi_{22} := \Phi_{111^\prime 1^\prime}$.  We further assume $\mu_{0}$, $\Phi_{22}$ are real\footnote{Im($C_{00}$) is assumed to be very large compared to $\Phi_{22}\prime/(2 \mathit{a}^{2} \Phi_{22})$ and Re($C_{00}$) to justify this approximation, while this condition mainly affects the differential equation for Im($\gamma$) only.} and use Einstein field equation in dyad form of the NP formalism \cite{NP1962}, and then an ordinary differential equation as \textit{RG-Friedmann equation} is obtained by such simple model. 
\begin{equation}
	\frac{\ddot{\mathit{a}}}{\mathit{a}}=-\frac{\mu _0 \mathcal{H}}{\mathit{a}}-\frac{\mathcal{H}_0^2 \Omega _M}{2 \mathit{a}^3}-\frac{1}{3} \mu _0^2 \mathcal{R}_M+2 \mathcal{H}^2
	\label{eq:RGFriedmann},
\end{equation}
where $\mathcal{H}$ is Hubble parameter, $\Omega_{M} := R_{m} \Omega_{B}, C_{0} := Re (C_{00})$, and $R_{m}$ is defined as following:
\begin{equation}
C_{0}= R_{m} \mu_{0},
\end{equation}
$R_{m}$ is the proportionality-constant for ordinary matter density parameter ($\Omega_{b}$) and apparent mass density parameter ($\Omega_{M}$). 
(It is applicable to the matter dominated era with mass density  $\sim \mathit{a}^{-3}$, although one can derive the equation for radiation dominated era in exactly the same way.)

It consists of matter and dark energy (DE) terms as we desire. Although it is not complicated, it is still not apparent to us how equations (\ref{eq:RGFriedmann}) can be solved analytically except for few simple scenarios.  Therefore we solve it numerically in this paper\footnote{The initial conditions are $\mathit{a}(t_{0})=1$ and $H(t_{0})=1$, and we set $t_{0}=1$.}

The DE part of Friedmann equation provides total 3 terms, a constant term, a dynamical term related to $\mathcal{H}/\mathit{a}$, as well as a positive dynamical term related to $\mathcal{H}^{2}$. Clearly the positive term is responsible for acceleration of expansion.

The Einstein equation in spinor formalism \cite{NP1962} is used to relate the curvature and energy momentum density spinor. Yet, the current theory of RG plays no role to explain this connection.

To find the equation of state of DE, $w(z)$, we use the usual definition as:
\begin{equation}
	DE := \Omega_{DE} H_{0} ^{2} \exp\left[3 \int^{1}_{0} \frac{1+w(x)}{1+x} dx\right]
	\label{eq:DEEOSDef}.
\end{equation}
The equation of state for the DE is found:
\small
\begin{multline}
  w(z)=(6  \mathcal{H}^2 (z)+\mu_0 (z+1)^2 \mathcal{H}'(z)\\
  -\mu _0^2  \mathcal{R}_M-2 (z+1) \mathcal{H}(z) (\mu_0+2  \mathcal{H}'(z)))/ \\
  (\mu _0^2  \mathcal{R}_M+3 \mathcal{H}(z) \left(\mu _0-2  \mathcal{H}(z)+\mu _0 z\right)).
	\label{eq:DEEOSDef2}
\end{multline}
\normalsize

In order to probe the constants of the cosmological model, we preliminarily use the DE-Matter equality: at $t\approx 0.7, z\approx 0.3$ (\textit{$H_{0}$ is set to 1 in this paper}).  It is found that one can set the RG-Friedmann equation to be an effective equation (applicable in DE, and matter dominated eras) for $\mu_{0} \approx 0$ limit.  The effective equation is:
\begin{equation}
  \frac{\ddot{\mathit{a}}}{\mathit{a}}=2 \mathcal{H}^2-\frac{\mathcal{H}_0^2 \Omega _M}{2 \mathit{a}^3}.
\end{equation}	

\begin{figure}[ht]
\includegraphics[width=\linewidth]{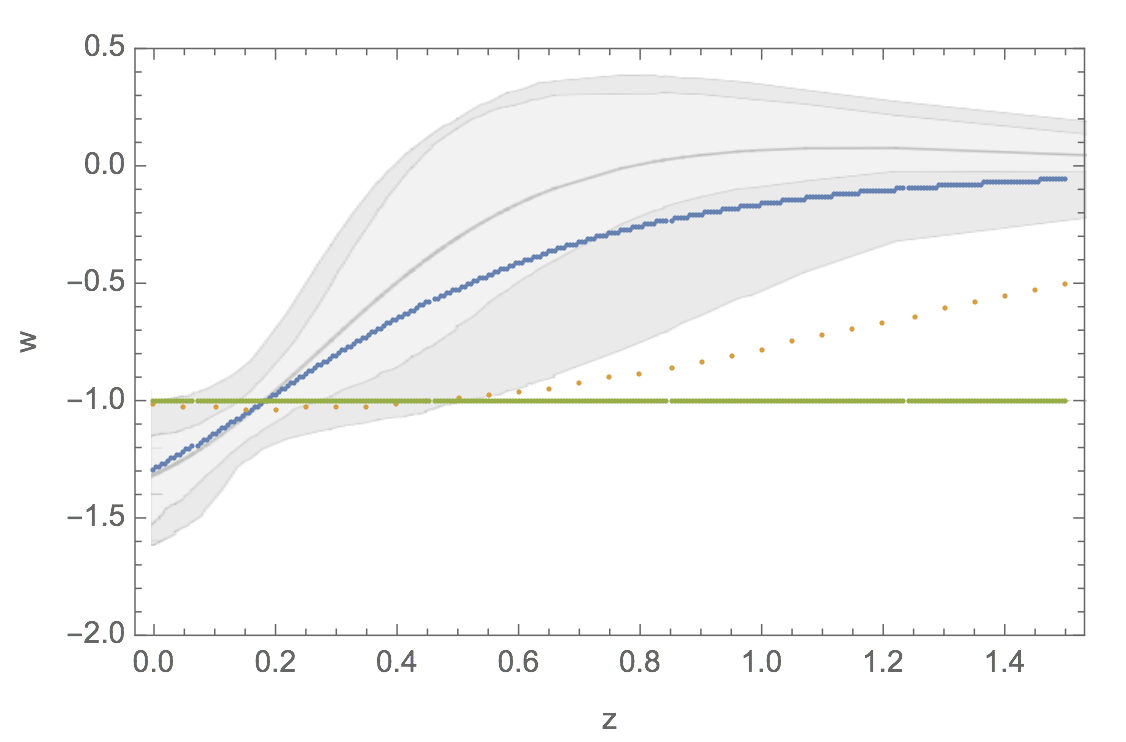}
\caption{\label{fig:RGDEw} Solid curved line:  the equation of state of effective DE model ($\Omega_{M} = 1.12$).  Dotted line: the equation of state of DE for the parameter-set $(\Omega_{M}$ $=0.156$, $\mu_{0}=-1.38)$.  The background plot is the fitting plot for polynomial form of DE based on data of SNIa(\cite{Li:2011sd}), the crossing point is obvious in low $z$ regime.  ($w=-1$ line is added as reference of cosmological constant)}
\end{figure}

With single parameter probe- $\Omega_{M}$ $\approx 1.12$, the predictions are:
\begin{center}
\textit{w(z) is near -1 in low z regime; w(z) provides a crossing to line w=-1; w(z) approaches 0 for high z regime}
\end{center}
[FIG.~\ref{fig:RGDEw}].  This is convincing comparing to the observational-best-fits from SNIa (see \cite{Li:2011sd} for a review).  Therefore, RG seems to be a good candidate for dynamical DE because of the fit from first principle considerations of the RG.

\begin{figure}[ht]
\includegraphics[width=\linewidth]{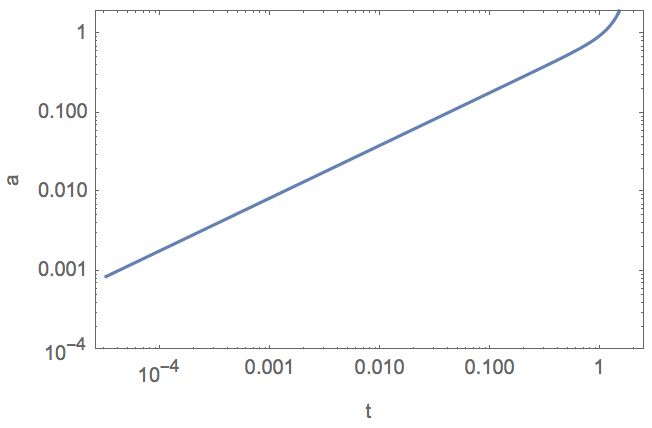}
\caption{\label{fig:RGDElogalogt} Log-Log plot of scale factor vs time ($t_{0}=1$) for effective DE model}
\end{figure}

\begin{figure}[ht]
\includegraphics[width=\linewidth]{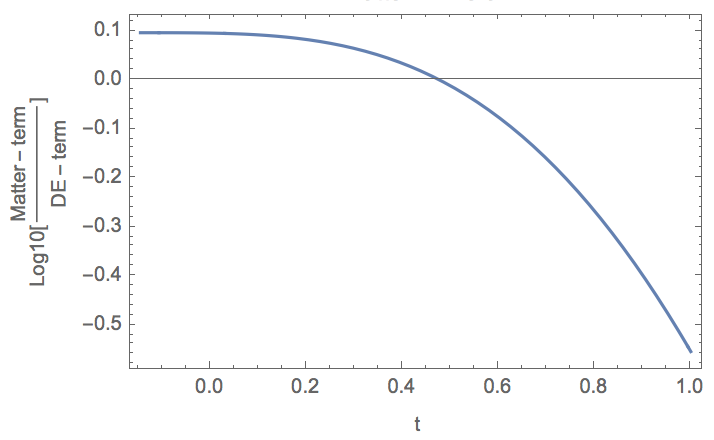}
\caption{\label{fig:RGDEMatDeEq} Log10 plot of ratio of matter term to DE term for effective DE model, it shows the shift of dominated term ($t_{0}=1$, age of universe is $\approx 1.15 H_{0}, H_{0}=1$)}
\end{figure}

Furthermore, the slope of expansion history in matter era is shown to be consistent with $\Lambda$CDM model [FIG.~\ref{fig:RGDElogalogt}].  It is not surprising, as $w(z)$ approaches zero as $z$ goes large that implies DE effect fades away as going back in time.  In addition, the plot of ratio of matter term vs DE term [FIG.~\ref{fig:RGDEMatDeEq}] is shown below for effective model ($\mu_{0} \approx 0$) to illustrate the consistency of accelerating expansion regime and the shift from matter dominated era to DE era.

Finally, beside the effective model of RG, there is a parameter-set $(\Omega_{M}$ $=0.156$, $\mu_{0}=-1.38)$ that the equation of state turns around $z \approx 0.2 $[FIG.~\ref{fig:RGDEw}].  In this case, $w(z)$ is very close to -1 in low $z$ regime while crossing from $w< -1$ to  $w>-1$ then approaches zero in high $z$.  However, for such parameters, RG predicts a much longer age of universe $\approx 2.38$.  Therefore, the validity is not obvious.

\section{COMMENTS AND POTENTIAL DEVELOPMENT}

By the simple physical concept of deformation on a Ratio, RG hypothesizes a simple entity to be fundamental mathematical structure of gravity.

The effective model with a simple form provides observation-consistent predictions yet only one parameter is needed.  It can be considered relatively simpler, compared to $\Lambda$CDM for matter and DE eras, since $\Lambda$CDM relies on two parameters for these two eras to be explained simultaneously. Therefore, the authors suggest RG proposes a non-phenomenological approach to explain the DE dynamics by the first principal of Ratio Gravity.  It is apparently justified because the parameters are linked directly to the Ratio context instead of arbitrary settings.

Since the main objective of this paper is to develop the RG framework for gravity, and establish its connection to the late-time universe, the current proposed parameters are some values chosen by hand inspired by observations. In the future, we hope to perform a more sophisticated scan of the parameter space of the theory with the most up-to-date data. 

To extend the model, one can consider more spin coefficients being non-zero, to describe a more complicated metric. Then, the related Bianchi constraints should have more equations to be solved.  The curvature matrices of $J$, $M$ have numbers of functions (e.g. the diagonal elements of $J$, $M$ matrices) to be reserved to extend to richer curvature structure. We hope to explore these possibilities in the future.

It is also possible to further extend the model from the Ratio context (rather than geometry). For example, the $C_{0}$ constant can be relaxed first, then more dynamic components in spin coefficients or curvature matrices should enter the dynamics.  We can also allow more degrees of freedom for matrix $Y$, such that more non-zero components of matrices are in place with the system to be solved.

The DE term drives the acceleration of the expansion so it is natural to ask: whether and how can the inflation era be explained from the same framework of RG?  More detailed studies are needed in order to study the early universe.

Finally, to explain the CMB fluctuations, we should investigate the primordial inhomogeneity as perturbations from RG prospective. We also leave the study of these fluctuations to future work.

\begin{acknowledgments}
JCHL would like to thank Michael Altman for introducing him to YW. We thank Henry Tye for discussions.
This work is supported in part by ECS Grant 26300316 and GRF Grant 16301917 from the Research Grants Council of Hong Kong.
\end{acknowledgments}


\bibliography{RGDE_correction_1_0}

\end{document}